\documentclass[11pt]{article}
\usepackage{moriond,epsfig}

\def\Journal#1#2#3#4{{#1} {\bf #2}, #3 (#4)}

\def\PLB{{\em Phys. Lett.}  B}

\def\ZPC{{\em Z. Phys.} C}
\def\JPG{{\em J. Phys.} G}

\def\jpsi{J/$\psi$~}
\def\psip{$\psi^\prime$~}

\def\be{\begin{equation}}
\def\ee{\end{equation}}
\def\bea{\begin{eqnarray}}
\def\eea{\end{eqnarray}}

\begin{document}
\vspace*{4cm}
\title{CHARMONIA PRODUCTION AT THE CERN/SPS}
\author{G.~Borges for the NA50 Collaboration\\
B.~Alessandro$^{10}$,
C.~Alexa$^{3}$,
R.~Arnaldi$^{10}$,
M.~Atayan$^{12}$,
S.~Beol\`e$^{10}$,
V.~Boldea$^{3}$,
P.~Bordalo$^{6,a}$,
G.~Borges$^{6}$,
C.~Castanier$^{2}$,
J.~Castor$^{2}$,
B.~Chaurand$^{9}$,
B.~Cheynis$^{11}$,
E.~Chiavassa$^{10}$,
C.~Cical\`o$^{4}$,
M.P.~Comets$^{8}$,
S.~Constantinescu$^{3}$,
P.~Cortese$^{1}$,
A.~De~Falco$^{4}$,
N.~De~Marco$^{10}$,
G.~Dellacasa$^{1}$,
A.~Devaux$^{2}$,
S.~Dita$^{3}$,
J.~Fargeix$^{2}$,
P.~Force$^{2}$,
M.~Gallio$^{10}$,
C.~Gerschel$^{8}$,
P.~Giubellino$^{10}$,
M.B.~Golubeva$^{7}$,
A.A.~Grigoryan$^{12}$,
S.~Grigoryan$^{12}$,
F.F.~Guber$^{7}$,
A.~Guichard$^{11}$,
H.~Gulkanyan$^{12}$,
M.~Idzik$^{10,b}$,
D.~Jouan$^{8}$,
T.L.~Karavicheva$^{7}$,
L.~Kluberg$^{9,5}$,
A.B.~Kurepin$^{7}$,
Y.~Le~Bornec$^{8}$,
C.~Louren\c co$^{5}$,
M.~Mac~Cormick$^{8}$,
A.~Marzari-Chiesa$^{10}$,
M.~Masera$^{10}$,
A.~Masoni$^{4}$,
M.~Monteno$^{10}$,
A.~Musso$^{10}$,
P.~Petiau$^{9}$,
A.~Piccotti$^{10}$,
J.R.~Pizzi$^{11}$,
F.~Prino$^{10}$,
G.~Puddu$^{4}$,
C.~Quintans$^{6}$,
L.~Ramello$^{1}$,
S.~Ramos$^{6,a}$,
L.~Riccati$^{10}$,
H.~Santos$^{6}$,
P.~Saturnini$^{2}$,
E.~Scomparin$^{10}$,
S.~Serci$^{4}$,
R.~Shahoyan$^{6,c}$,
F.~Sigaudo$^{10}$,
M.~Sitta$^{1}$,
P.~Sonderegger$^{5,a}$,
X.~Tarrago$^{8}$,
N.S.~Topilskaya$^{7}$,
G.L.~Usai$^{4}$,
E.~Vercellin$^{10}$,
L.~Villatte$^{8}$,
N.~Willis$^{8}$,
T.~Wu$^{8}$}

\address{
$^{~1}$ Universit\`a del Piemonte Orientale, Alessandria and INFN-Torino, Italy
$^{~2}$ LPC, Univ. Blaise Pascal and CNRS-IN2P3, Aubi\`ere, France
$^{~3}$ IFA, Bucharest, Romania
$^{~4}$ Universit\`a di Cagliari/INFN, Cagliari, Italy
$^{~5}$ CERN, Geneva, Switzerland
$^{~6}$ LIP, Lisbon, Portugal
$^{~7}$ INR, Moscow, Russia
$^{~8}$ IPN, Univ. de Paris-Sud and CNRS-IN2P3, Orsay, France
$^{~9}$ Laboratoire Leprince-Ringuet,  Ecole Polytechnique and 
CNRS-IN2P3,  Palaiseau,  France
$^{10}$ Universit\`a di Torino/INFN, Torino, Italy~
$^{11}$ IPN, Univ. Claude Bernard Lyon-I and CNRS-IN2P3, Villeurbanne, France
$^{12}$ YerPhI, Yerevan, Armenia
\\
a) also at IST, Universidade T\'ecnica de Lisboa, Lisbon, Portugal~
b) also at AGH University of Science and Technology, Faculty of Physics and 
Applied Computer Science, Krakow, Poland
c) on leave of absence of YerPhI, Yerevan, Armenia}

\maketitle\abstracts{We present the final results of experiment NA50 on charmonia 
production in Pb-Pb interactions at 158\,A\,GeV. A strong increasing 
suppression is observed with increasing centrality, for both the \jpsi and \psip 
resonances. We also present new developments regarding the \jpsi and \psip normal 
nuclear absorption determinations deduced from proton-nucleus data only. Their 
comparison with Pb-Pb results allows us to conclude that the \jpsi anomalous suppression 
sets in at mid-centralities while the S-U results show a reasonable agreement with the 
\mbox{p-A} behavior. The \psip suffers a significantly stronger suppression already 
in S-U interactions, which continuously increases in the Pb-Pb system, and is completely
incompatible with the expected behavior deduced from p-A collisions.}

\section{Introduction}
\label{sec:intro}
The aim of the NA50 experiment is to search for a transition from 
normal hadronic nuclear matter to a quark-gluon plasma (QGP) phase of deconfined 
quarks and gluons. Theory has predicted 
that \jpsi suppression in ultrarelativistic heavy ion collisions would unambiguously sign   
such a transition. We present here the latest and final results on charmonia 
production in Pb-Pb interactions at 158\,A\,GeV  from data collected by the experiment 
at the CERN SPS. The apparatus, optimized for the detection and study of muon pairs 
produced in heavy ion collisions, is well adapted to identify and measure, through its 
leptonic decay, the production of charmonium in such collisions. 
The study has been extended to proton-nucleus reactions 
which provide the appropriate reference for any possible new feature 
specific of heavy ion collisions.  

Charmonia resonances are measured by the NA50 spectrometer through their decay into a pair 
of muons in the 2.92$< y_{\rm lab} <$3.92 and $|{\rm cos}(\theta_{\rm CS})|<$0.5 kinematical 
window. The muon spectrometer, made of an air core toroidal magnet in-between two sets of 
four multiwire proportional chambers and two trigger hodoscopes, allows to trigger the 
experiment on the detection of two muons simultaneously produced in the target region and to  
measure their trajectories and momenta. A hadron absorber placed 
downstream from the target is used as a muon filter for the spectrometer. 
In the NA50 Pb-Pb configuration of the setup, the detector also provides,  
on an event-by-event basis, the centrality 
of the interaction as measured by three different devices: an electromagnetic calorimeter 
(EMC), a forward hadronic calorimeter (ZDC) and a multiplicity detector (MD).
Charmonia production in Pb-Pb collisions can thus be studied as a function of three different centrality 
estimators: $E_{\rm T}$, the neutral transverse energy; $E_{\rm ZDC}$, the beam spectators energy;
and $N_{\rm ch}$, the charged particles multiplicity. 
The different sources which contribute to the dimuon samples collected in the experiment   
are extracted from a simultaneous fit to the invariant opposite-sign dimuon mass spectra, 
obtained for each centrality bin, where all known physical sources are included. Drell-Yan 
dimuons, the only contribution for the very high mass part of the spectrum, are used as 
a reference since their 
production cross-section scales linearly with the number of nucleon-nucleon collisions, as 
experimentally shown \cite{Abr97}. Results are normally 
presented as ratios of charmonia to Drell-Yan cross-sections, with the advantage 
that, since Drell-Yan is measured using the same invariant mass spectrum as the \jpsi and \psip 
resonances, most of the systematic errors cancel out in these ratios. The drawback of this
procedure is the relatively small Drell-Yan sample which becomes the major source of 
statistical uncertainties.

\section{\jpsi and \psip nuclear absorption: The normal  references}
\label{sec:nnac}
\jpsi normal nuclear absorption is deduced from the analysis of a large set of 
proton-nucleus data, collected at 450\,GeV and 400\,GeV and covering a 
wide range of nuclear sizes. A simultaneous Glauber analysis of the ratios
\mbox{$B_{\mu\mu} \sigma({\rm J}/\psi)/\sigma({\rm DY}_{2.9-4.5})$} from three different 
NA50 p-A data sets, together with the pp and pd results obtained by the NA51 Collaboration \cite{NA51}, 
lead, for the normalization and absorption cross-section of the \jpsi 
resonance through nuclear matter, to fitted values of 57.5$\pm$0.8 (at 450\,GeV) and 
$\sigma_{\rm abs}({\rm J}/\psi)$=4.2$\pm$0.4\,mb, 
respectively. These values, when used as inputs in a Glauber model, allow to compute 
the expected \jpsi production in a Pb-Pb system. Since there are no data allowing to perform
a $\sigma_{\rm abs}({\rm J}/\psi)$ systematic study as a function of $\sqrt{s}$, we 
assume that the previous $\sigma_{\rm abs}({\rm J}/\psi)$ value holds at 158\,GeV. 
The normalization of the absorption curve at the Pb-Pb energy is obtained by rescaling the \jpsi and 
Drell-Yan contributions separately. The \jpsi scaling factor is obtained through the analysis of the
\mbox{$B_{\mu\mu} \sigma({\rm J}/\psi)$/A} values, for which there are data available at 200\,GeV, 
obtained by the NA38 \cite{NA38} and NA3 \cite{NA3} experiments. From a simultaneous Glauber fit at 
the different energies (see left panel of figure \ref{fig:1}), we measure 
$\sigma_{\rm abs}({\rm J}/\psi)$=4.1$\pm$0.4\,mb, a value in excellent agreement with the one 
extracted using the ratios \mbox{$B_{\mu\mu} \sigma({\rm J}/\psi)/\sigma({\rm DY}_{2.9-4.5})$}. 
We extract 0.319$\pm$0.025 as the factor which scales down the \jpsi production cross-section 
from 450 to 200\,GeV \cite{Bor04}. 
The additional scaling needed to go from 200 to 158\,GeV (0.737$\pm$0.006) is 
deduced using a phenomenological description of the $\sqrt{s}$ dependence of the 
cross-section~\cite{Schuler}, $\rm \sigma_{\psi} = \sigma_0 (1-M_\psi / \sqrt{s})^n$. 
All the available experimental data have been fitted, leading to n=12.8$\pm$0.3. The 
$x_{\rm F}$ correction also needed (1.020$\pm$0.013) is computed~\cite{Schuler} 
based on a $x_{\rm F}$ distribution parametrized as a function of $\sqrt{s}$. 
The Drell-Yan process is scaled down from 450 to 158\,GeV 
(0.387$\pm$0.010) using a LO calculation with the same parton distribution 
functions (GRV 94 LO) as the one used in the Pb-Pb and p-A data analysis. 
The \psip normal nuclear absorption is deduced from the ratios 
\mbox{$B^\prime_{\mu\mu} \sigma(\psi^\prime)/\sigma({\rm DY}_{2.9-4.5})$}
which, fitted with the same Glauber approach as used for the \jpsi, give 1.08$\pm$0.05
and 7.6$\pm$1.1\,mb, for the normalization and \psip absorption cross-section, respectively.
The corresponding absorption curve at 158\,GeV is obtained using the same scaling factors
as for the \jpsi.

\section{\jpsi anomalous suppression} 
\begin{figure}
\centering
\psfig{figure=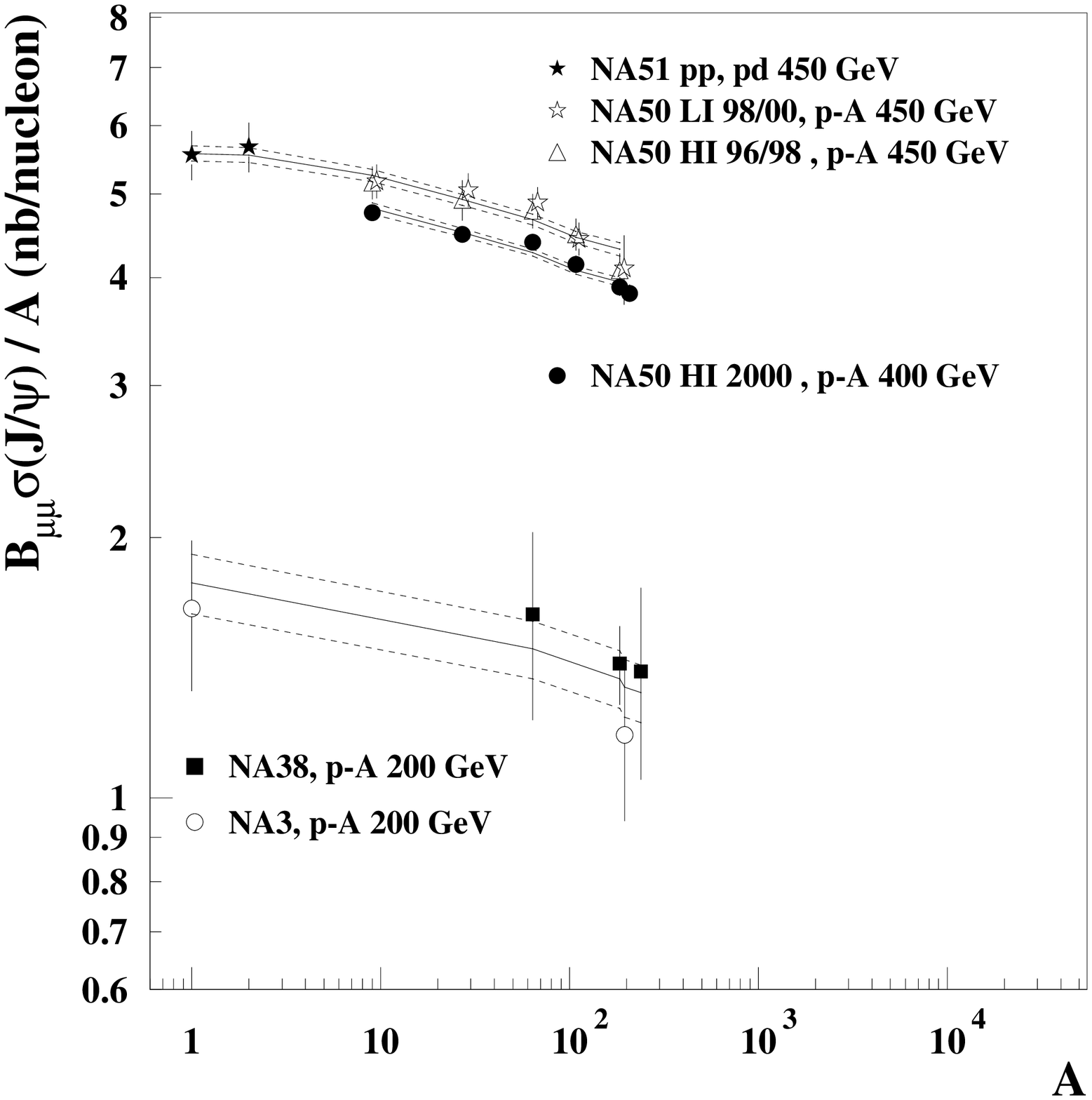,height=0.21\textheight}
\psfig{figure=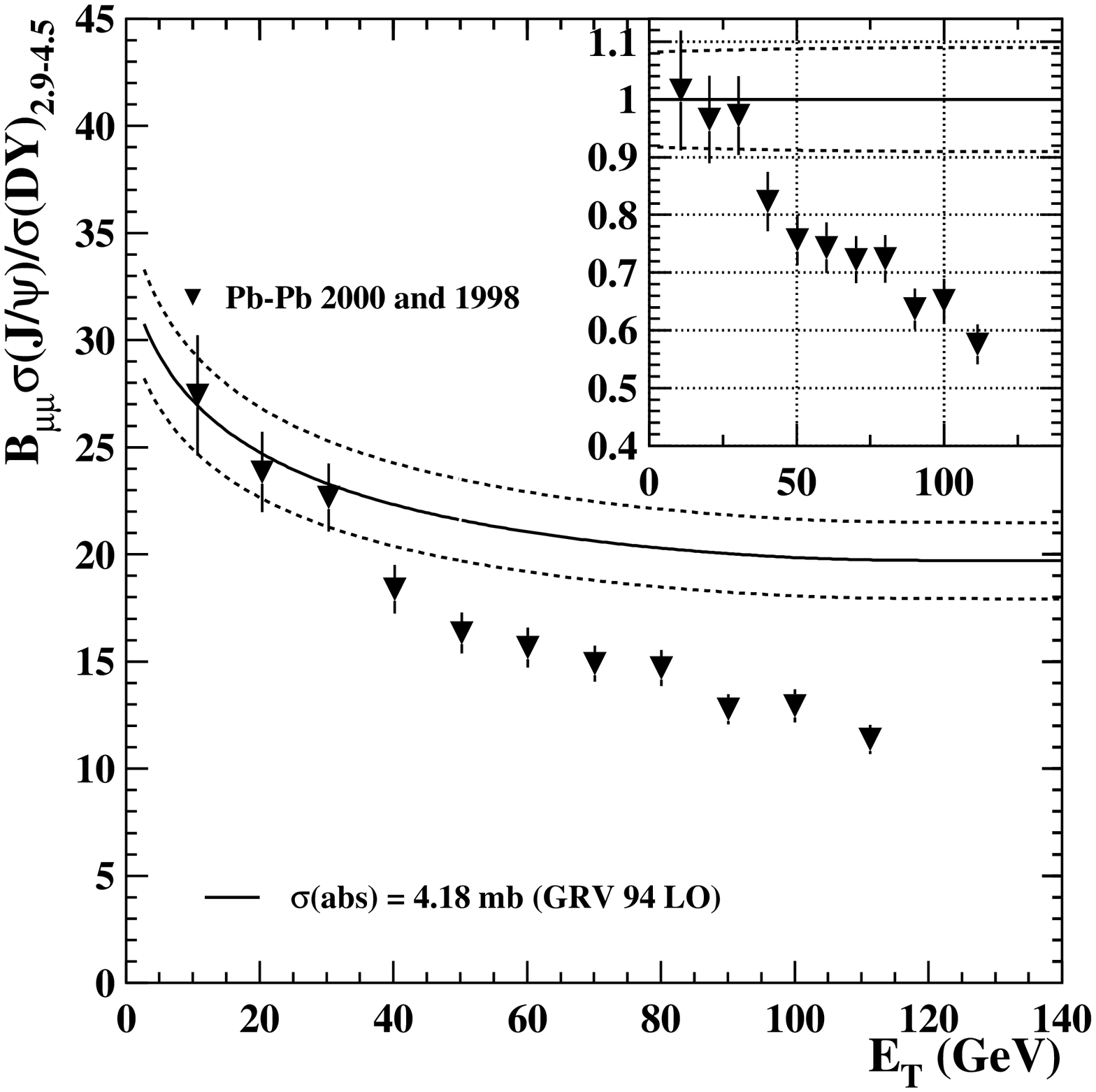,height=0.21\textheight}
\vspace{-0.25cm}
\caption{Simultaneous Glauber fit using the ratio \mbox{$B_{\mu\mu} \sigma({\rm J}/\psi)$/A} 
to extract the \jpsi rescaling factors (left panel). The ratios J/$\psi$/DY in Pb-Pb are 
compared with the nuclear absorption curve as a function of $E_{\rm T }$ (right panel).}
\label{fig:1}
\end{figure}

The ratio $B_{\mu\mu} \sigma({\rm J}/\psi)/\sigma({\rm DY}_{2.9-4.5})$ measured in Pb-Pb
is now compared
with the nuclear absorption curve obtained following the procedures explained in the previous
section. The results are presented as a function of 
centrality  and averaged, for each separate centrality region, from data collected in years 
1998 and 2000. 
Indeed, a good statistical compatibility between both data sets has been obtained, 
after reanalyzing the already published 1998 data with the same criteria as used for the 2000 analysis. 
A comparison
as a function of any of the three centrality estimators shows the same J/$\psi$/DY pattern: 
a clear departure from the normal nuclear absorption curve at mid-centrality values and a
suppression which then increases with increasing centrality (see right panel of figure \ref{fig:1}).
All available data (p-A, S-U and Pb-Pb) can be plotted as a function of L, the average path length
crossed by the $\rm c\bar{c}$ pair through nuclear matter, which makes 
possible a direct comparison from pp up to the most central Pb-Pb region. We observe that the S-U 
$B_{\mu\mu} \sigma({\rm J}/\psi)/\sigma({\rm DY}_{2.9-4.5})$ results show a reasonable 
compatibility with other lighter systems while the Pb-Pb central measurements are anomalously suppressed
(see left panel of figure~\ref{fig:2}).

\section{\psip suppression}
\begin{figure}
\centering
\psfig{figure=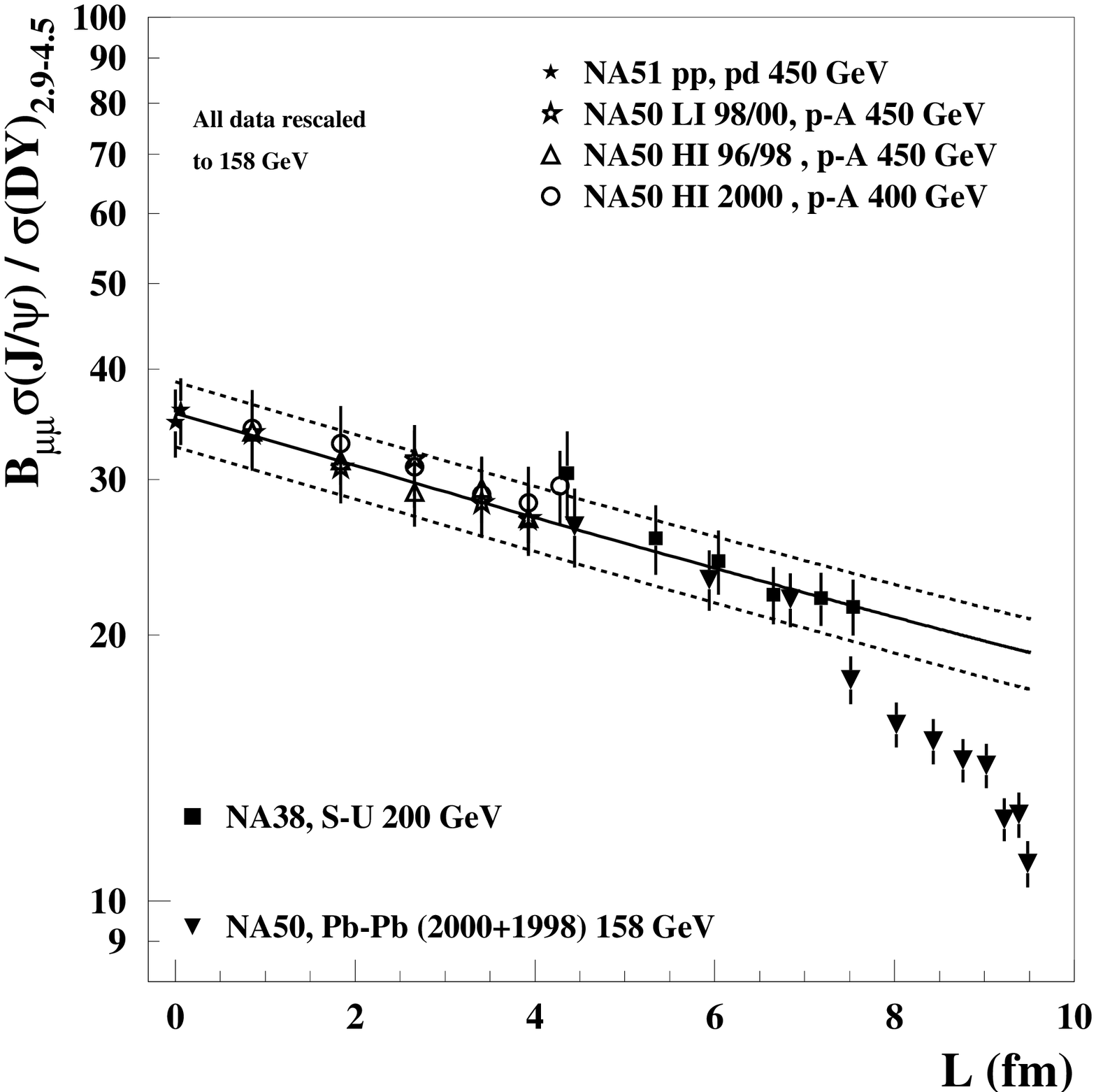,height=0.21\textheight}
\psfig{figure=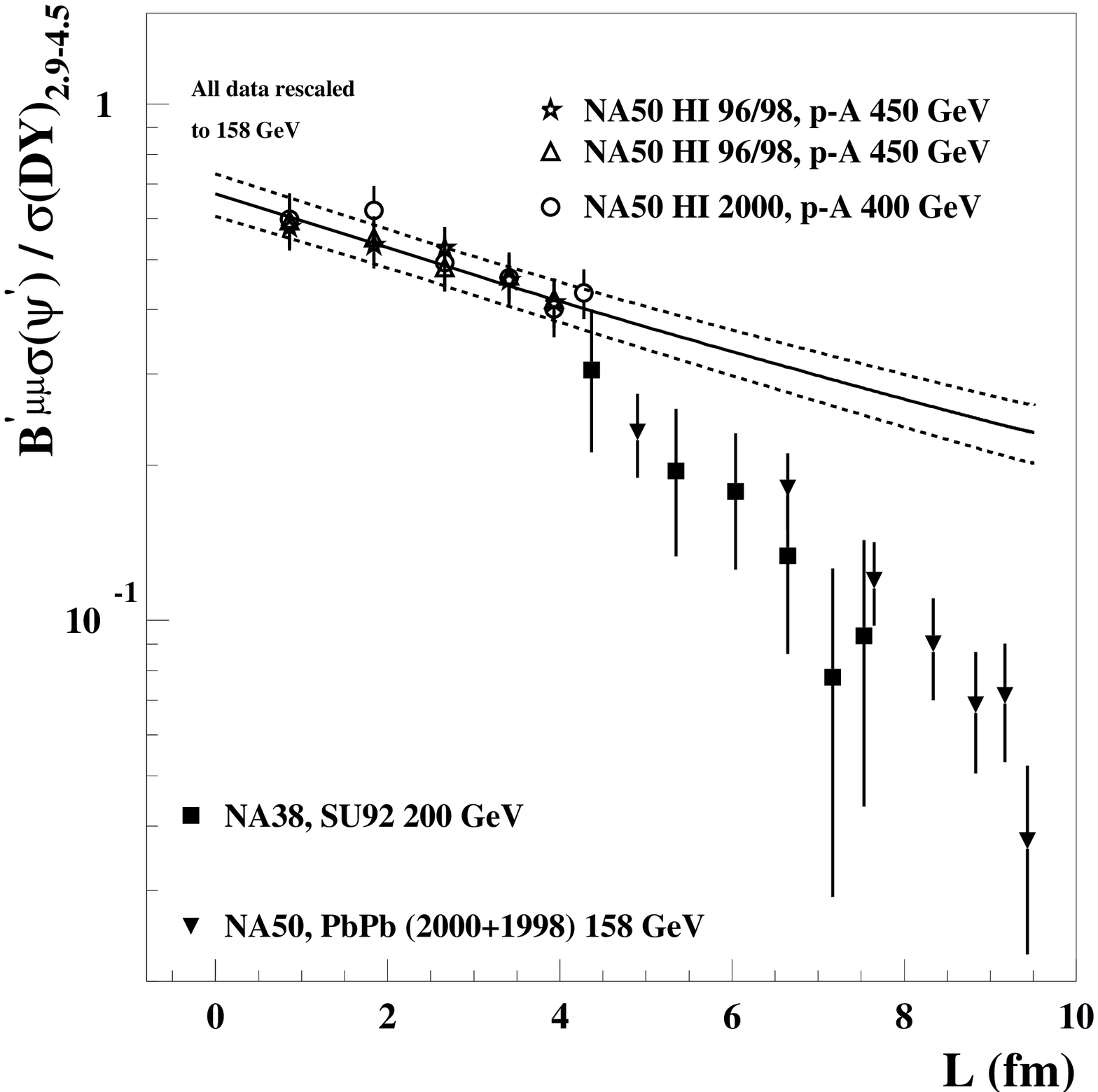,height=0.21\textheight}
\psfig{figure=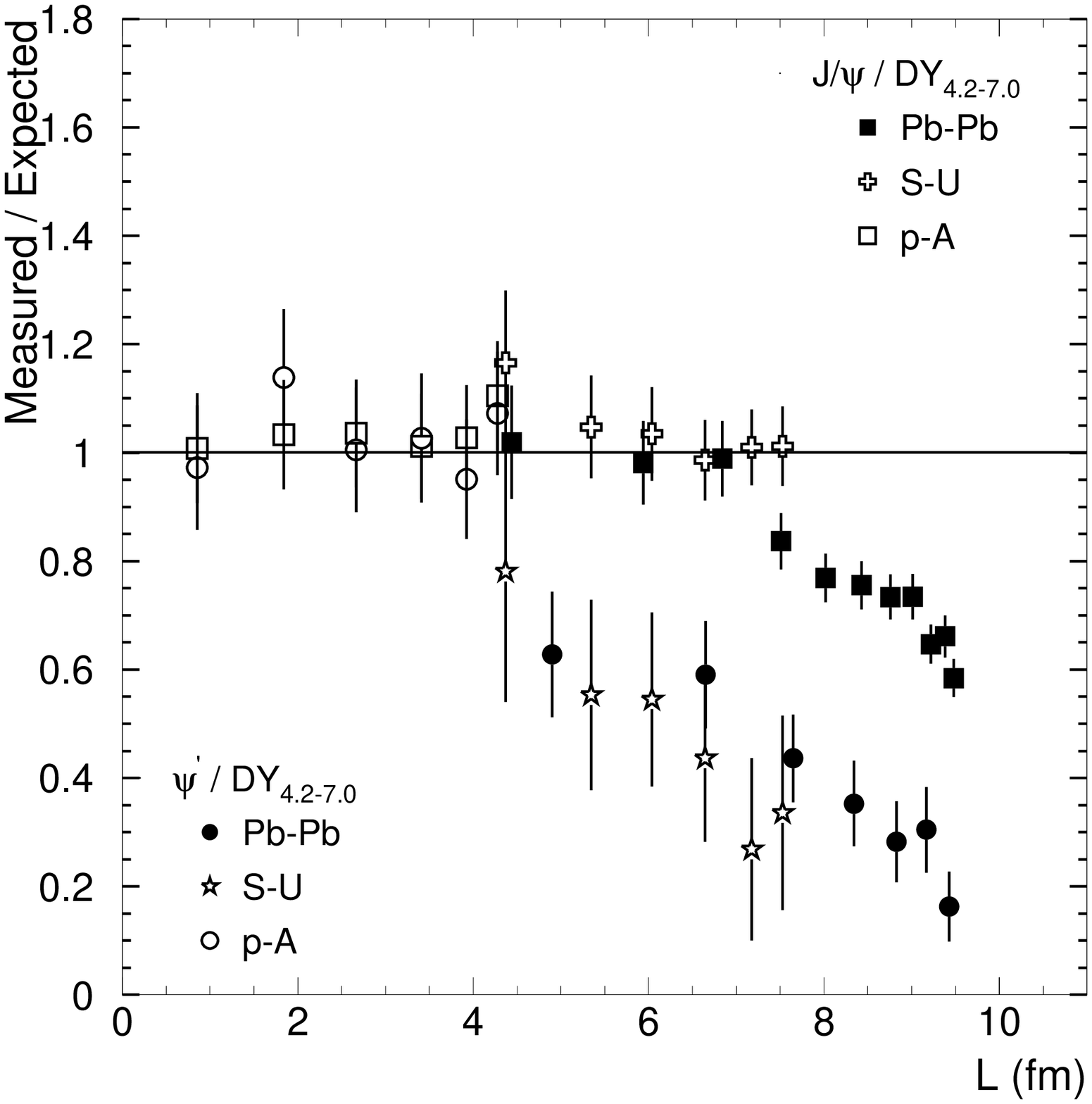,height=0.21\textheight}
\vspace{-0.25cm}
\caption{The ratios J/$\psi$/DY  as a function of L (left panel) and $\psi^\prime$/DY
as a function of L (center panel). The same ratios
divided by the expected values deduced from p-A collisions (right panel).}
\label{fig:2}
\end{figure}

The \psip results are extracted from the same opposite sign dimuon mass spectra as the \jpsi 
\cite{San04}. However, this is a more delicate analysis due to the small 
\psip cross-section 
and to the overlap of several dimuons sources in the \psip mass region, which can strongly influence 
the results of the fit. The difficulties increase with centrality since the \psip is more and more 
suppressed with respect to those other sources.
 Careful systematic studies 
have been performed. We observe a strong suppression of the $\psi^\prime$ in Pb-Pb 
collisions. If the direct ratio between the two charmonia states is used to compare 
both suppression patterns, 
we observe a factor $\sim$\,2.5 higher suppression for the \psip resonance, between peripheral 
and central collisions. Once again, the L variable is used to put together the ratios $\psi^\prime$/DY 
for various systems 
as a function of centrality, which leads to the conclusion that a strong \psip suppression 
(with respect to p-A) sets-in already in S-U collisions and that the same continuous pattern 
is observed for Pb-Pb interactions (see center panel of figure~\ref{fig:2}).

\section{Conclusions}
We have presented final results from NA50 on charmonia production in Pb-Pb collisions
using the Drell-Yan process 
as reference. The comparison of ion induced \jpsi production yields with the expected 
behavior inferred from proton-nucleus collisions shows a reasonable compatibi\-li\-ty 
of S-U and peripheral Pb-Pb measurements with the normal nuclear absorption curve. \jpsi 
anomalous suppression sets-in at mid-centrality and systematically increases for smaller 
impact parameter interactions. The \psip production in Pb-Pb follows the S-U pattern as a function 
of centrality but is completely incompatible with the behaviour expected from proton-nucleus reactions.
Divi\-ding the measured charmonia to DY ratios by the expected normal nuclear absorption curves,  
we conclude that the
\psip suppression sets-in earlier with respect to the \jpsi suppression and presents, for the
same centrality values, lower surviving probabilities (see right panel of figure~\ref{fig:2}).
 
\section*{Acknowledgments}
This work was partially supported by the Funda\c{c}\~ao para a Ci\^encia e Tecnologia, Portugal.

\end{document}